\documentclass[twocolumn,showpacs,preprintnumbers,amsmath,amssymb,APS]{revtex4}

\usepackage{dcolumn}
\usepackage{bm}
\usepackage[spanish,english]{babel}
\usepackage{amsfonts}
\usepackage{amssymb}
\usepackage{graphicx}
\usepackage[latin1]{inputenc}
\newcommand{\R}{\mathcal{R}}

\begin{document}

\title{ {\bf Bouncing Cosmologies in Palatini $f(R)$ Gravity}}

\author{Carlos Barragán}
\affiliation{ \footnotesize Departamento de Física Teórica, Universidad Autónoma de Madrid,  28049 Madrid, Spain}

\author{Gonzalo J. Olmo}\email{olmo@iem.cfmac.csic.es}
\affiliation{ \footnotesize Instituto de Estructura de la Materia, CSIC, Serrano 121, 28006 Madrid, Spain}

\author{Hèlios Sanchis-Alepuz}
\affiliation{ \footnotesize Departamento de Física Teórica and IFIC, Centro Mixto Universidad de Valencia-CSIC.  Facultad de Física, Universidad de Valencia, Burjassot-46100, Valencia, Spain \\ and \\ Fachbereich Theoretische Physik, Institut für Physik, Karl-Franzens-Universität Graz,
Universitätsplatz 5, A-8010 Graz, Austria}

\date{June 12$^{th}$, 2009}

\begin{abstract}
We consider the early time cosmology of $f(R)$ theories in Palatini formalism and study the conditions that guarantee the existence of homogeneous and isotropic models that avoid the Big Bang singularity. We show that for such models the Big Bang singularity can be replaced by a cosmic bounce without violating any energy condition. In fact, the bounce is possible even for pressureless dust. We give a  characterization of such models and discuss their dynamics in the region near the bounce.  We also find that power-law lagrangians with a finite number of terms may lead to non-singular universes, which contrasts with the infinite-series Palatini $f(R)$ lagrangian that one needs to fully capture the effective dynamics of Loop Quantum Cosmology. We argue that these models could also avoid the formation of singularities during stellar gravitational collapse. 
\end{abstract}


\maketitle

\section{Introduction}

Under the dynamical laws of General Relativity (GR), the standard FLRW cosmology becomes singular at the origin of time. The energy density, expansion rate, and other geometrical invariants diverge as the volume of the Universe goes to zero. The Big Bang singularity seems to be an unavoidable and disturbing aspect of the currently established cosmological model \cite{Weinberg08} which, to the eyes of many, only a full quantum theory of gravity could resolve. Though a completely satisfactory quantum theory of gravitation is not yet available, some interesting progress has been made in the last years in the field of quantum cosmology. The application of non-perturbative canonical quantization techniques, developed within the framework of Loop Quantum Gravity (LQG) \cite{LQG}, to FLRW universes has led to a very active research program known as Loop Quantum Cosmology (LQC) \cite{LQC}. The chief accomplishment of this line of research is the dynamical resolution of the Big Bang singularity, which is replaced by a cosmic bounce at an energy density of the order of the Planck density.  Though the dynamics of LQC is governed by difference equations, which is reminiscent of the discrete structure of spacetime characteristic of LQG, an effective continuum description in terms of differential equations is possible. These equations represent modified Friedmann dynamics such that the relation between matter/energy and curvature is different from that found in GR. As a result, at energies close to the Planck scale curvature scalars reach an extremum that cannot be dynamically surpassed, whereas at low matter/energy densities the equations smoothly recover those of GR. Thus, in LQC the matter generates curvature but, unlike in GR, {\it it cannot break the spacetime}. The singularity is, thus, resolved by introducing no new degrees of freedom in the theory, just by changing the way matter curves the geometry. \\

	These properties of LQC, namely the ability to curve spacetime differently than in GR but without introducing new degrees of freedom, can also be found in modified theories of gravity of the $f(R)$ type in Palatini formalism \cite{Palatini}(which are radically different from the more standard $f(R)$ theories in metric formalism [see, for instance, \cite{Olmo05} for a comparison]). This fact was first pointed out in \cite{Olmo-Singh-09} and was used there to see if a Palatini $f(R)$ lagrangian with the same cosmological dynamics as LQC could be found \footnote{The same problem has been considered more recently using an $f(R)$ theory in metric formalism \cite{Sot-09}. However, the perturbative approach followed there implies that that result can only be trusted at low curvatures and, therefore, nothing can be stated about the high curvature regime. The lagrangian found in \cite{Olmo-Singh-09}, however, is valid even in the region of the bounce.}. The answer was in the affirmative, and a unique solution to the problem was found. Though the lagrangian was obtained using numerical methods, two approximate analytical functions which fit very well the numerical curve at low and high curvatures, respectively, were also given. A remarkable property of those functions is that they can be expressed as infinite power series of the curvature, which puts forward the fact that in non-perturbative LQC quantum geometry effects are non-local.\\
	
To the light of the results of \cite{Olmo-Singh-09} one may wonder how generic are bouncing cosmologies in Palatini $f(R)$. Can we somehow characterize the lagrangians that avoid the Big Bang singularity? It is also reasonable to ask if such lagrangians are necessarily tied to infinite power series, which are associated with the non-local nature of curvature found in LQC. These questions and others will be explicitly addressed in this paper. We will see that the conditions for a lagrangian to generate a cosmic bounce are limited and also that finite power series lagrangians can produce non-singular cosmologies. In fact, we show that the simplest model that one can imagine, namely $f(R)=R+R^2/R_P$ has such a property, where $R_P$ is the Planck scale curvature. This model, in turn, leads to cosmic bounce even for presureless dust, and that makes it particularly interesting for the analysis of simple gravitational collapse models. We will comment on this aspect and on the extensions with respect to the FLRW model needed to correctly address the stellar collapse model à la Oppenheimer-Snyder. \\

The paper is organized as follows. In section \ref{sec:FLRW} we discuss the FLRW dynamics of Palatini theories and identify the conditions for bouncing cosmology to exist. Then we consider a simple power-law lagrangian which satisfies the bouncing conditions and study its dynamics. In \ref{sec:BH} we consider a stellar collapse model following the classical treatment of Oppenheimer-Snyder and show that such a model is insufficient to guarantee the matching with the external Schwarzschild solution at all times. We conclude with a summary and some conclusions.

\section{Bouncing Cosmologies \label{sec:FLRW}}

The action that defines Palatini $f(R)$ theories is of the form 
\begin{equation}\label{eq:Pal-Action}
S[{g},\Gamma ,\psi_m]=\frac{1}{2\kappa^2}\int d^4
x\sqrt{-{g}}f({R})+S_m[{g}_{\mu \nu},\psi_m]
\end{equation}
Here $f({R})$ is a function of ${R}\equiv{g}^{\mu \nu }R_{\mu \nu }(\Gamma )$, with $R_{\mu \nu }(\Gamma )$ given by
$R_{\mu\nu}(\Gamma )=-\partial_{\mu}
\Gamma^{\lambda}_{\lambda\nu}+\partial_{\lambda}
\Gamma^{\lambda}_{\mu\nu}+\Gamma^{\lambda}_{\mu\rho}\Gamma^{\rho}_{\nu\lambda}-\Gamma^{\lambda}_{\nu\rho}\Gamma^{\rho}_{\mu\lambda}$
where $\Gamma^\lambda _{\mu \nu }$ is an independent connection (not the Levi-Civita connection of $g_{\mu \nu }$). The matter action $S_m$ depends on the matter fields $\psi_m$ and the metric $g_{\mu\nu}$. Variation of the action with respect to metric and connection leads to the following equations
\begin{eqnarray}
f_R R_{\mu\nu}(\Gamma)-\frac{f}{2}g_{\mu\nu} &=& \kappa^2 T_{\mu\nu}\label{eq:met-var1} \ , \\
\nabla_{\beta}\left[\sqrt{-g}f_R g^{\mu\nu}\right]&=&0  \label{eq:con-f_R} \ , 
\end{eqnarray}
where $f_R$ denotes $df/dR$. Using the trace of (\ref{eq:met-var1}), 
\begin{equation}\label{eq:trace}
Rf_R-2f=\kappa^2T\ ,
\end{equation}
one obtains an algebraic relation between $R$ and $T$, which will be denoted by $R=\R(T)$ and  generalizes the GR relation $R=-\kappa^2T$. Note that for a given lagrangian one may find multiple solutions $\R_i(T)$ of (\ref{eq:trace}) whose physical validity must be studied model by model. It could also happen that (\ref{eq:trace}) had no real solutions, which must be regarded as unphysical. In physically acceptable models, the algebraic relation $R=\R(T)$ allows us to see the $f_R$ term of (\ref{eq:con-f_R}) as a function of $T$. The connection can then be solved as the Levi-Civita connection of an auxiliary metric  $h_{\mu\nu}$ conformally related with the physical metric, $h_{\mu\nu}=f_R g_{\mu\nu}$. Using this relation, we can rewrite (\ref{eq:met-var1}) as follows
\begin{eqnarray}\label{eq:neweinstein}
G_{\mu \nu}(g) &=& \frac{\kappa^2}{f_R} T_{\mu \nu} - \frac{\R f_R - f}{2 f_R} g_{\mu \nu} + \frac{1}{f_R}\left(\nabla_\mu \nabla_\nu f_R-g_{\mu \nu} \Box f_R\right)  \nonumber\\ 
&& - \frac{3}{2 f_R^2} \left(\partial_\mu f_R \partial_\nu f_R - \frac{1}{2} g_{\mu \nu} (\partial f_R)^2\right) \ ,
\end{eqnarray}
where the $\R, f_R,$ and $f_{RR}$ terms on the right hand side are functions of $T$, all covariant derivatives are referred to the metric $g_{\mu\nu}$, and the Einstein tensor on the left hand side is also referred to the metric $g_{\mu\nu}$. \\

When particularized to FLRW cosmologies for matter with constant equation of state $P=w\rho$, (\ref{eq:neweinstein}) becomes
\begin{eqnarray}
3H^2&=&\frac{f_R\left(\kappa^2\rho+V/2\right)}{\left(f_R+\frac{3}{2}\Lambda_1\right)^2}-\frac{3K}{a^2} \label{eq:H} \\
(2f_R+3\Lambda_1)\dot H&=&-3H^2\left[f_R(2- 3\Lambda_2)+(2-3w)\Lambda_1\right]\nonumber \\
&+& [\kappa^2\rho(1-w)+V] -f_R\left(\frac{4K}{a^2}\right)\label{eq:Ray}
\end{eqnarray}
where $H\equiv \dot a/a$, $K$ is the curvature of the spatial homogeneous sections, and we have defined
\begin{eqnarray}\label{eq:FLRW}
V&\equiv& R f_R-f \\
\Lambda_1 &\equiv& f_{RR}(\alpha_w \rho R_T)\\
\Lambda_2 &\equiv& \frac{f_{RRR}}{Rf_{RR}-f_R} (\alpha_w \rho R_T)^2 \\
R_T &= & \frac{\kappa^2}{R f_{RR}-f_R} \\
\alpha_w &\equiv & (1+w)(1-3w)
\end{eqnarray}
To obtain (\ref{eq:H}) and (\ref{eq:Ray}) we have used the conservation law $\dot \rho=-3H(1+w)\rho$ and the trace  equation (\ref{eq:con-f_R}).  Note that since $R=R(\rho)$, the functions $f_R, f_{RR}, V, \Lambda_1, \Lambda_2$ and $R_T$ appearing in (\ref{eq:H}) and (\ref{eq:Ray}) are all functions of $\rho$. \\

One can verify that GR is recovered in the limit of linear lagrangian, $f_{RR}\to 0$. When the non-linearities are important, equations (\ref{eq:H}) and (\ref{eq:Ray}) represent modified Friedmann equations in which the cosmic expansion is altered by the new role played by the various $\rho-$dependent terms. This fact was used in \cite{Olmo-Singh-09} to find an $f(R)$ lagrangian that could reproduce the effective dynamics of  LQC. In the next subsection we summarize those results. 

\subsection{Loop Quantum Cosmology as an $f(R)$}

The effective dynamics of LQC is characterized by the following Friedmann equation (with $K=0$)
\begin{equation}\label{eq:LQC}
3H^2=\kappa^2\rho\left(1-\frac{\rho}{\rho_{crit}}\right)
\end{equation}
where $\rho_{crit}\approx 0.41\rho_{Planck}$ is the maximum dynamically accessible energy density of the theory. An $f(R)$ lagrangian reproducing this equation (and also the corresponding equation for $\ddot a$) can be found by identifying the right hand side of (\ref{eq:LQC}) with that of (\ref{eq:H}) and replacing $\rho$ by the rule given in (\ref{eq:trace}), with $T=2\rho$ for a massless scalar. One then obtains a second order differential equation for the function $f(R)$ which can be numerically solved subject to certain boundary conditions, namely, agree with GR at low curvatures and give the right acceleration $\ddot a$ at the position of the bounce. The numerical curve was very well fitted by the following ansatz (see Fig. 1)
\begin{equation}\label{eq:ansatz}
f(R)=-\int dR \tanh\left(\frac{5}{103}\ln \left[\frac{R}{12R_c}\right]^2\right)
\end{equation}
where $R_c\equiv \kappa^2\rho_{crit}$. It is important to note that for this isotropic, spatially flat model, the numerical results indicate that the bounce occurs when $f_R=0$, which here corresponds to $R=-12R_c$. One can readily verify from (\ref{eq:H}) that $f_R=0$ is one of the possible conditions for a vanishing $H$ when $K=0$. Next we discuss the conditions for a bounce to exist in the cases with $K=-1,0,+1$. 
\begin{figure}
\includegraphics[width=0.45\textwidth]{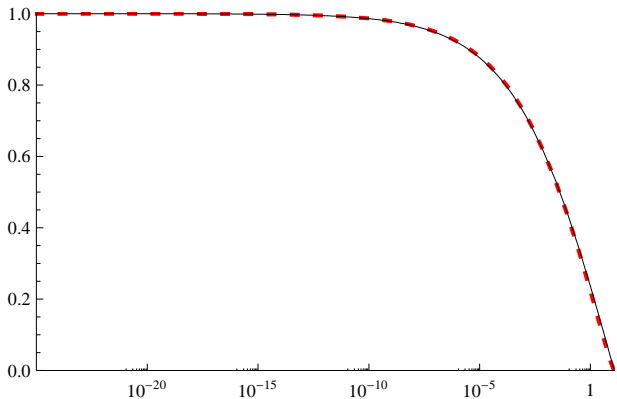}
\caption{Comparison of the ansatz $f_R$ given in (\ref{eq:ansatz}) (solid line) with the numerical curve obtained in \cite{Olmo-Singh-09} (dashed line). The plot represents $df/dR$ versus $|R|$. \label{fig:LQC}}
\end{figure}

\subsection{Characterization of the Bounce.}

We now consider the conditions for getting $H=0$ which follow from (\ref{eq:H}). We assume that our lagrangian is almost linear at low curvatures, i.e., $\lim_{R\to 0} f_R\to 1 $ and may depart from unity at high curvatures near the bounce. We do not consider models with infrared corrections due to their pathological physical behavior (see the discussion in \cite{Olmo08} and references therein). For convenience, we define $\Delta\equiv\kappa^2\rho+V/2$, which must be positive at low curvatures (assuming $K/a^2$ negligible) because $H^2$ is always positive. In this generic discussion we also assume that the term $f_R+\frac{3}{2}\Lambda_1$ does not vanish, which would imply a divergence of $H^2$ and, therefore, a Big Bang singularity. That possibility will be discussed later, when particular models are considered. We also note that the particular equation of state $w=1/3$ must be excluded from the analysis because that particular case implies that $T=0$ and, therefore, the dynamics of the theory is exactly that of GR plus an effective cosmological constant, which is of no interest in our discussion. 

\subsubsection{$K=0$ case.}
In the spatially flat case there are two possible conditions for the bounce:
\begin{itemize}
\item{Case A: $f_R=0$ at some $R\sim R_P$.} 

This condition is the one that we find in LQC.  From (\ref{eq:trace}), we see that if $f_R=0$ at some high density $\rho_B$, where the subindex $B$ denotes {\it bounce}, then $f(R_{B})=(1-3w)\kappa^2\rho_{B}/2$. This means that at $\rho_B$ we find $\Delta_B=3(1+w)\kappa^2\rho_B/4$. Since $\Delta$ cannot be negative within the interval $\rho\in [0,\rho_B]$ (because $H^2\ge 0$), then the type A bounce can only occur for sources with $w>-1$.  \\
Let us now focus on the region near the bounce. Since in the limit $f_R\to 0$ the density reaches its maximum value $\rho_B$, (\ref{eq:Ray}) can be approximated by 
\begin{equation}\label{eq:H-approx}
\dot H\approx -(2-3w)H^2+\alpha^2
\end{equation}
where $\alpha^2\equiv \frac{\kappa^2\rho_B(1-3w)+V_B}{3\Lambda_1^B}$ is a positive constant, which follows from the condition for a minimum at $f_R=0$. The solution of (\ref{eq:H-approx}) is given by 
\begin{eqnarray}
H^2(t)_{w<\frac{2}{3}}&=& \frac{\alpha^2}{2-3w}\tanh^2[\alpha\sqrt{2-3w}(t-t_B)]  \\
H^2(t)_{w>\frac{2}{3}}&=& \frac{\alpha^2}{3w-2}\tan^2[\alpha\sqrt{3w-2}(t-t_B)] 
\end{eqnarray}
where $t_B$ is an integration constant that sets the instant at which the bounce occurs. The corresponding expansion factor is given by
\begin{eqnarray}
a(t)_{w<\frac{2}{3}}&=& a_B (\cosh[\alpha\sqrt{2-3w}(t-t_B)])^\frac{1}{2-3w}  \\
a(t)_{w>\frac{2}{3}}&=& a_B (\cos[\alpha\sqrt{3w-2}(t-t_B)])^\frac{1}{2-3w} 
\end{eqnarray}
Note that these results are valid for any lagrangian $f(R)$ near the bounce caused by the condition $f_R=0$.

\item{Case B: $\Delta=0$ at some $R\sim R_P$.}

If starting with $f_R>0$ and $\Delta>0$ we assume that the bounce occurs when $\Delta=0$, which requires $f(R_B)=R_B f_{R_B} (1+3w)/(3+3w)$, then equation (\ref{eq:trace}) implies that $R_B f_{R_B}=-3(1+w)\kappa^2\rho_B$. Restricting again our attention to the cases $w>-1$, we see that when $\Delta=0$ then $R_B f_{R_B}<0$. If this happens keeping $f_{R_B}>0$ then this condition could be physically possible. Otherwise, the condition $f_R=0$ would have  been already satisfied and we would be in case A. To determine the sign of $f_{R_B}$, we focus on the sign of $R$, which we assume to be a monotonous function of $\rho$. Assuming that at low curvatures the GR relation $R=(1-3w)\kappa^2\rho$ is a good approximation, then $R>0$ if $-1<w<1/3$ and $R<0$ if $w>1/3$ \footnote{Note that the case $w=1/3$ corresponds to pure radiation and turns the field equations into those of $GR$ plus an effective cosmological constant.}. Therefore, if $w>1/3$ it is possible to find lagrangians $f(R)$ for which cosmic bounce occurs before $f_R=0$. 

\end{itemize}

\subsubsection{Cases $K\neq 0$.}

In this cases, the discussion is not as clean as in the $K=0$ case, and only a qualitative description is possible. For $K>0$, we see that the right hand side of (\ref{eq:H}) contains a negative term. Therefore, the bounce will occur before reaching the conditions $f_R=0$ or $\Delta=0$ of the $K=0$ case, i.e., at the bounce $f_R>0$ and $\Delta>0$. On the contrary, if we choose $K<0$, then the curvature term on the right hand side of (\ref{eq:H}) is always positive and the bounce can only occur if the first term becomes sufficiently negative, which would require either $f_R>0,\Delta<0$ or $f_R<0,\Delta>0$. The latter condition is problematic because it implies a change of sign in the function $f_R$, which plays a crucial role in the resolution of the connection field equation (\ref{eq:con-f_R}). In fact, the connection is solved as the Levi-Civita connection of an auxilary metric which is conformally related with the physical metric, i.e., $h_{\mu\nu}=f_R g_{\mu\nu}$. The vanishing of $f_R$ then raises doubts on the physical validity of such solutions. However, we will see later that the numerical evolution of such models does not present any pathological behavior (at least for the models chosen) when the function $f_R$ crosses the zero. To understand why this solution is also physically valid and well behaved, we are forced to interpret the $f(R)$ theory as a scalar-tensor theory. In fact, it is well known that $f(R)$ theories in metric and Palatini formalism are dynamically equivalent to the $w=0$ and $w=-3/2$ Brans-Dicke theories, respectively, with the identifications $\phi\equiv f_R$ and $V(\phi)=R f_R-f$. In the Palatini case, $w=-3/2$, the scalar field is non-dynamical and can be algebraically expressed as a function of the trace $T$, i.e., $\phi=\phi(T)$. In this formulation of the theory, there is no reason, in principle, to forbid a change of sign in $\phi$, because one works from the very beginning with $g_{\mu\nu}$ and there is no need to assume that the conformal transformation  $h_{\mu\nu}=\phi g_{\mu\nu}$ must always be well defined. One may also wonder if the dynamics is well defined at the instant when $\phi$ becomes zero. This question is pertinent because the field equations can be written as $\phi G_{\mu\nu}=\kappa^2T_{\mu\nu}+$ ``other terms''. It turns out that the ``other terms'' that appear on the right hand side, also contain second derivatives of the expansion factor which are finite when $\phi=0$. Therefore, at least for FLRW cosmologies, the dynamics is well defined even when $\phi=0$. Therefore, these solutions should also be seen as physically possible. In any case, since the current establishment is that $K\gtrsim 0$ we will not care any more about the possibility $K<0$.

\section{Toy model: $f(R)=R+\frac{R^2}{R_P}$}

In this section we consider a simple polynomial model with quadratic deviations from GR, $f(R)=R+R^2/R_P$. Note that $R_P$ is not restricted to be positive. This model is particularly interesting because it leads to the familiar relation $\R=-\kappa^2T$, which will make easier the discussion and the formulas. We will first focus on the case $K=0$ to illustrate how the existence of a cosmic bounce depends on the choice of equation of state and the sign of $R_P$. Then we consider the case $K>0$, which can be used as a first approach in the modeling of stellar gravitational collapse à la Oppenheimer-Snyder. We will show that periodic bounces and expansions occur in this model but that a complete description of gravitational collapse must go beyond this simple FLRW model for the stellar interior (see section \ref{sec:BH}). We will also show that cosmic bounce exists for $K<0$.\\

\subsection{Bouncing Universe with $K=0$.\label{sec:K0}}

As we mentioned before, the bounce will occur whenever $f_R=0$ or $\Delta=0$ before the denominator of the right hand side of (\ref{eq:H}) can have have the chance to vanish, which would imply a singularity. For the model chosen here, (\ref{eq:H}) turns into
\begin{equation}\label{eq:H-R2}
H^2=\frac{R}{(1-3w)}\frac{\left(1+\frac{2\R}{R_P}\right)\left(1+\frac{1-3w}{2}\frac{\R}{R_P}\right)}{\left[1-(1+3w)\frac{\R}{R_P}\right]^2}
\end{equation}
The condition $f_R=0$ occurs when $\R_B=-R_P/2$, which translates into  $\kappa^2\rho_B=-R_P/2(1-3w)$. From this it follows that if $R_P>0$ then the bounce can only occur for sources with $w>1/3$. If $w<1/3$ then, we need to flip the sign in front of $R_P$ to guarantee the bounce, which is equivalent to taking the lagrangian $f(R)=R-R^2/|R_P|$. In both cases, the denominator does not vanish before or at $R=\R_B$ if $w>-1$. This bouncing mechanism is thus possible for sources with $w>-1$ (recall that the radiation dominated case $w=1/3$ is excluded in general). \\

The condition $\Delta=0$, occurs when $\R_B^{(\Delta)}=-2R_P/(1-3w)$, which tranlates into $\kappa^2\rho_B^{(\Delta)}=-2R_P/(1-3w)^2$. This forces $R_P$ to be negative and implies that only $f(R)=R-R^2/|R_P|$ could produce a $\Delta-$induced bounce. For this condition to be a valid bouncing mechanism, we should have $f_R>0$ and the denominator should not vanish anywhere neither before nor at the limiting curvature $\R_B^{(\Delta)}$. However, it is easy to see that $f_R>0$ can only be satisfied before or at $\R_B^{(\Delta)}$ if $w<-1$ or $w>1/3$ while, at the same time, to avoid the vanishing of the denominator before or at $\R_B^{(\Delta)}$ we need $-1<w<1/3$. Since these two conditions cannot be satisfied together, we conclude that in these models there cannot be a $\Delta-$induced bounce. \\

\begin{figure}
\includegraphics[width=0.55\textwidth]{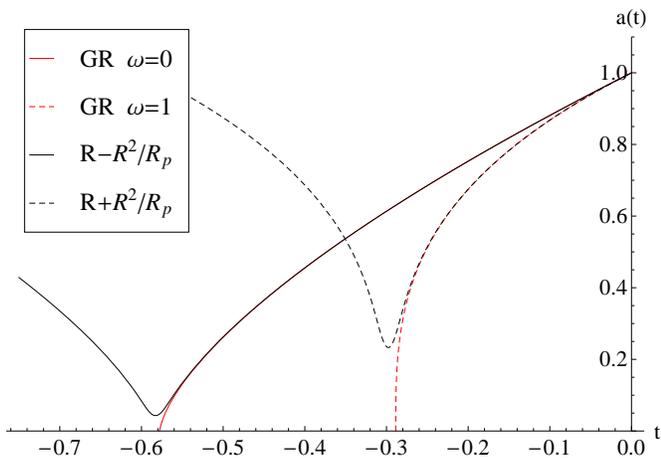}
\caption{Bouncing cosmologies with $K=0$ compared with the GR solution (red lines). The continuous black curve represents the case $w=0$, corresponding to $f(R)=R-R^2/|R_P|$. The dashed line represents the case $w=1$ of the theory $f(R)=R+R^2/|R_P|$. The initial density is the same for the four curves. \label{fig:K0}}
\end{figure}

In summary, for the model $f(R)=R+R^2/|R_P|$ the bounce is possible only for sources satisfying $w>1/3$ whereas for the model $f(R)=R-R^2/|R_P|$ the bounce is possible only for $-1<w<1/3$. In both cases, the bounce occurs when $f_R=0$. All other cases lead to universes with Big Bang singularities. See Fig.\ref{fig:K0} for examples with $K=0$.\\

\subsection{ Cyclic Cosmology with $K>0$.\label{sec:K+}}

\begin{figure}
\includegraphics[width=0.55\textwidth]{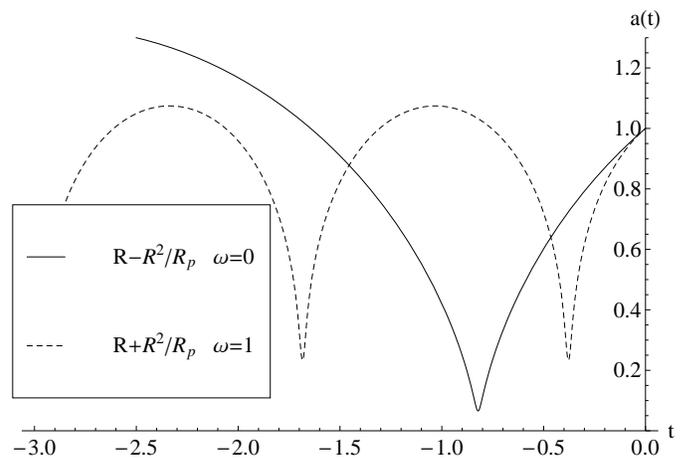}
\caption{Cyclic cosmologies with $K>0$. The continuous black curve represents the case $w=0$, corresponding to $f(R)=R-R^2/|R_P|$. The dashed line represents the case $w=1$ of the theory $f(R)=R+R^2/|R_P|$. The initial density is the same as in Fig. \ref{fig:K0}. \label{fig:K+}}
\end{figure}

The condition $K>0$ implies that the Universe has a finite maximum radius. Since the modified dynamics of our model also sets a minimum radius for certain configurations, the resulting non-singular cosmologies are cyclic. We illustrate this cyclic behavior in Fig. \ref{fig:K+}, where the same initial density and models as in the case $K=0$ are shown. 

\subsection{ Bouncing Universe with $K<0$.\label{sec:K-}}

\begin{figure}
\includegraphics[width=0.55\textwidth]{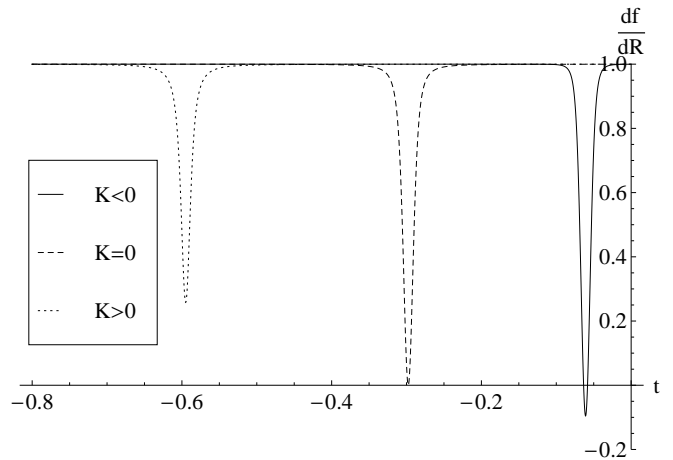}
\caption{Evolution of $f_R$ for $K=0$ (dashed line), $K>0$ (dotted line), and $K<0$ (solid line). The initial density is the same as in Fig. \ref{fig:K0}. \label{fig:K-}}
\end{figure}

The condition $K<0$ implies that the bounce occurs for negative values of $f_R$. In Fig. \ref{fig:K-} we show how the sign of $f_R$ can change depending on the choice of $K$. It is clear that the numerical curve is smooth and that nothing suggests a breakdown of the dynamics when $f_R=0$ is crossed. 

\section{Gravitational Collapse}\label{sec:BH}

The last stages of a collapsing massive star were first investigated by Oppenheimer and Snyder \cite{O-S39} (see also \cite{F-NS05} for a pedagogical discussion). They simplified the complexity of such  astrophysical  scenario by assuming that the interior of the star could be seen as an homogeneous and isotropic matter distribution. In this manner, the interior geometry can be described by means of a FLRW metric, whereas the exterior geometry takes the form of the Schwarzschild solution. If the collapsing star has a finite initial radius, then the interior metric must have positive spatial curvature, $K>0$. Demanding continuity of the metric components across the surface, one finds a relation between $K$, the mass, and the initial radius of the star. This simple model showed that for an observer comoving with the fluid, the radius of the star shrinks to zero and the density goes to infinity in a finite proper time. The conclusion was that the fate of sufficiently massive stars is the formation of a black hole. \\ 

A simplified analysis such as that carried out by Oppenheimer and Snyder, in which the matter profile is discontinuous at the surface of the star, is not (strictly speaking) possible in general for $f(R)$ Palatini theories. The reason is that in these theories, the matter must satisfy certain differentiability conditions which are not needed in the case of GR. This can be understood by noting that the field equations (\ref{eq:neweinstein}) for the metric in Palatini $f(R)$ contain terms of the form $\nabla_\mu\nabla_\nu f_R$, which turn into first and second derivatives of the trace $T$. For this reason, a discontinuous matter profile would generate divergences \footnote{Though a discontinuous matter profile is technically possible in GR, we must admit that matter at sufficiently short scales does require a continuous description in terms of fields. Therefore, the smoothness requirement of the matter profiles in Palatini theories should not be viewed as a deffect of the theory.} on the right hand side of the metric field equations (note that other type of divergences have been detected in spherically symmetric stellar solutions for some equations of state. Nonetheless, for the models considered here they could never be physically realized. See the discussion in \cite{BSM,Olmo08b} for details.). Therefore, to correctly study a stellar collapse process in Palatini $f(R)$ theories, one is (strictly speaking) forced to consider an inhomogeneous matter profile that smoothly interpolates between the interior distribution of matter and the external vacuum. Nonetheless, for the class of models that we are considering, these difficulties can be relaxed for some time and some qualitative results can be extracted.\\

\indent Explicit evaluation of the derivatives  $\nabla_\mu f_R$,  for $f(R)=R+R^2/R_P$, shows that those terms are strongly suppressed by $1/R_P$ factors. In fact, in order to get $|\nabla_\mu f_R|^2 \sim |\nabla_\mu \rho/\rho_P|^2$, where $R_P=\kappa^2\rho_P$, on the same order of magnitude as $\kappa^ 2\rho$ (the usual Einstein's equations dominant term), the matter density profile should change by order of the Planck density on scales of the order of the Planck length. This means that for a realistic initial configuration of macroscopic size, the interior can be well approximated as homogeneous and isotropic everywhere except at the outermost regions of the star, where the density profile must go to zero smoothly. If in this inhomogeneous region the length scale of variation of $T$ is much larger than the Planck scale $l_P\sim 1/\sqrt{|R_P|}$, then the radial derivatives $\nabla_r f_R$ can be neglected and global internal homogeneity is a good approximation. This would justify the use of the Oppenheimer-Snyder model to study the first stages of the stellar collapse in our theory. However, this approximate homogeneity cannot last for ever. As the star collapses, inhomogeneities must grow faster on the outer regions than in the inner ones because radial density gradients are larger out there. This should affect the rate at which the inner and outer layers of the star collapse. One thus expects that the innermost regions, where gradients are weaker, should evolve in a way similar to that found in the FLRW $K>0$ cosmologies of Fig.\ref{fig:K+}. This fact suggests that the singularity will be avoided. The outermost regions, on the contrary, will feel the effect of the density gradients and very little can be guessed about their fate. A complete description of the matching of these layers of the star with the external Schwarzschild solution would thus require detailed numerical analysis. The relevant question to answer now is whether the surface of the star manages to cross the Schwarzschild radius in a finite proper time or not, i.e., if an event horizon is formed or not. These issues will be studied elsewhere \cite{BdR-O-SA}.  \\

\section{Summary and Discussion}

In this work we have shown that Palatini $f(R)$ theories have unusual properties that make them specially interesting for addressing strong gravity phenomena such as the early Universe and stellar collapse processes. The simplest lagrangians with high curvature corrections that one can consider, namely $f(R)=R\pm R^2/|R_P|$,  lead to non-singular universes at high curvatures and seem compatible with non-singular collapsing stars. Repulsive gravitational forces arise in these scenarios induced by the time and/or radial derivatives of the matter-energy density profile. No new degrees of freedom in the gravitational side or in the matter side (exotic sources) are needed to get such repulsive gravitational forces. This miracle is possible due to the entanglement existing between matter and geometry induced by the independent connection used to construct the action. When the matter-energy density and pressure reach a certain high-energy scale, spacetime is curved differently than in GR and singularities are avoided, which contrasts with other known mechanisms to avoid cosmic singularities \cite{Novello.etal}. At low energies, as expected, the dynamics is identical to that of GR. We thus have theories, $f(R)=R\pm R^2/R_P$, with the same low energy behavior as GR but without its (classical) high-energy drawbacks. \\

Quadratic lagrangians of the form $R+\alpha R_{\mu\nu}R^{\mu\nu}+\beta R^2$ have been thoroughly studied in the literature in the usual variational metric formalism. It is well known from the works of Starobinsky, Schmidt, and Müller \cite{Staro80,MuSch85,StaSch87,MuSchSta88} that all (vacuum) solutions of such models which recover at late times power-law FLRW universes do not avoid the initial Big Bang singularity. The only nonsingular solutions of those models describe, rather than a bouncing Universe, a universe starting from a de Sitter state. When in those models $\alpha$ is set to zero, $f(R)=R+\beta R^2$, bouncing solutions can be found, though they are incompatible with a power law expanding Universe at late times (see the discussion in \cite{Novello.etal} and references therein).  The Palatini $R\pm R^2/|R_P|$ models discussed here do not suffer from such problems. The properties of the generalized models $R+\alpha R_{\mu\nu}R^{\mu\nu}+\beta R^2$ will be studied elsewhere. \\

Our results suggest that Palatini theories might play a relevant role in the phenomenology of gravitation at high energies. It has already been shown that these theories are able to reproduce the dynamics of isotropic LQC. The formulation of that theory in terms of holonomies and triads and the non-existence of a connection operator seems to be related with the fact that an infinite series in $R$ is needed to reproduce that dynamics à la Palatini. The fact that finite polynomial lagrangians can also lead to non-singular universes indicates that non-local dynamics is not essential to remove singularities in second-order theories. On the other hand, the assumption of independence between metric and connection in the variational process is essential to get second order equations for the metric. Since full LQG has the same number of dynamical fields as GR and Palatini $f(R)$, it is thus reasonable to expect that effective descriptions of that theory could come in the form of Palatini theories. Unfortunately, we are still very far from having under control the dynamical aspects of LQG, which makes impossible the search for effective equations. Nonetheless, it would be very positive if we could find further relations between Palatini theories and other aspects of the quantum gravity phenomenology, besides that of bouncing cosmologies. A better understanding of these theories in scenarios with less symmetries and with more general families of Palatini lagrangians could help us deepen into such aspects. That will be the subject of future studies. 

\acknowledgements

C.B. thanks  Universidad Autónoma de Madrid for financial support. G.J.O. thanks MICINN for a Juan de la Cierva contract and the Departamento de Física Teórica \& IFIC  of the University of Valencia - CSIC  for their hospitality during the elaboration of this work. The research of G.J.O. has also been partially supported by grant FIS2005-05736-C03-03.

\end{document}